\documentclass[aps,pra,
twocolumn,showpacs,psfig]{revtex4}

\usepackage{graphicx}
\usepackage{amsmath}
\usepackage{amssymb}

\begin{document}


\title{On hydrodynamic flow of expanding Bose-Einstein condensates}
\author{V.A. Brazhnyi$^{1}$, A.M. Kamchatnov$^{1,2}$, and V.V. Konotop$^{1}$}
\affiliation{$^{1}$Centro de F\'{\i}sica Te\'orica e Computacional, Universidade de Lisboa,
Complexo Interdisciplinar, Av. Prof. Gama Pinto 2, Lisboa 1649-003, Portugal }
\affiliation{$^{2}$Institute of Spectroscopy, Russian Academy of Sciences, Troitsk 142190,
Moscow Region, Russia}

\begin{abstract}
We study expansion of quasi-one-dimensional Bose-Einstein condensate (BEC)
after switching off the confining harmonic potential. Exact solution of
dynamical equations is obtained in framework of the hydrodynamic approximation
and it is compared with the direct numerical simulation of the full problem
showing excellent agreement at realistic values of physical parameters. We
analyze the maximum of the current density and estimate the velocity of
expansion. The results of the 1D analysis provides also qualitative
understanding of some properties of BEC expansion observed in experiments.

\end{abstract}

\maketitle

Many current experiments with Bose-Einstein condensates (BEC) include regime
of a free expansion of the gas which initially was (magnetically or optically)
trapped by a confining potential \cite{Stamper,Mewes}. The respective phase of
the BEC evolution is of practical interest since much information about
coherent matter waves is experimentally obtained at this stage (say, by
absorption imaging), and also because of importance of a free BEC flow in a
context of recently proposed new devices like, for example, an atomic laser
\cite{laser}. At ultra-low temperatures, a trapped BEC is well described by
the three-dimensional (3D) Gross-Pitaevskii (GP) equation which in the absence
of a trap potential is known as a nonlinear Schr\"odinger (NLS) equation. In a
number of specific cases it can be reduced to a quasi-one-dimensional one.
More specifically, this happens when one can neglect (due to some reasons)
interaction between transverse modes. In the absence of a trap the problem
thus reduces to an integrable model, 1D NLS equation, which is very well
studied fundamental model of the nonlinear physics.

The problem of theoretical description of evolution of BEC confined by
harmonic potential with varying parameters has been addressed in several
papers \cite{cast}. In the experiment \cite{bongs} the realization of BEC
expansion in quasi-1D waveguide has been reported and again excellent
agreement has been found with theoretical predictions at long enough values of
time of evolution. These results have been confirmed by numerical solution of
corresponding dynamical equations for quasi-1D case \cite{cast,ps02}. In the
present communication we give complete analytical treatment of this problem.
We consider only BECs with positive scattering lengths. In terms of the NLS
equation, the situation we are dealing with corresponds to the evolution of an
initially localized pulse with the initial profile corresponding to the ground
state of a BEC confined in a parabolic potential, i.e. to the defocusing NLS
equation with zero boundary conditions at infinity. As it is known, such
problem does not have soliton solutions and independently of number of
particles the condensate will spread out and its density will tend to zero
with time. The formal analytical description of the respective solution at $t
\to\infty$ was obtained rather long ago \cite{manakov}. However, for practical
purposes of analysis of experimental data it is desirable to have a
description of the condensate evolution during the initial stages of time,
too. It turns out that such evolution also admits rather complete analytical
description in a number of cases, and, in particular, when the initial density
distribution is smooth enough, or in terms of BEC in the limit of large number
of atoms which leads to the so-called Thomas-Fermi (TF) approximation. Then
the hydrodynamic approach allows one to describe analytically evolution of the
density and velocity fields of the condensate. The evolution of BEC at initial
stages appears to be quite rich. For example, as it has been recently shown in
\cite{SSS02}, it may display wave breaking for some specific initial
distributions of the BEC density.

In the present report we describe the free expansion of a BEC initially
confined in a harmonic potential. On the base of the hydrodynamic approach we
find the time dependence of the density and velocity distributions and
calculate such characteristics of the gas flow as the value and space and time
coordinates of the maximum of current, velocity of the ``edge'' points of the
condensate, as well as asymptotic density and velocity distributions. Also we
show that simple analytical estimates of the 1D problem allow one to give
qualitative understanding of the phenomena observed in experiments with
effectively 2D BEC.


Let us start with the 3D GP equation for the order parameter $\psi\equiv
\psi(\mathbf{r},t)$,
\begin{equation}
\label{GPE}i\hbar\frac{\partial\psi}{\partial t}=-\frac{\hbar^{2}}{2m}%
\Delta\psi+V_{trap}(\mathbf{r})\psi+g_{0}|\psi|^{2}\psi\, ,
\end{equation}
where we use the standard notation: $g_{0}=4\pi\hbar^{2}a_{s}/m$, $a_{s}$
being the $s$-wave scattering length, which is considered positive, $m$ being
the atomic mass; $V_{trap}(\mathbf{r})$ is a trap potential. Considering the
case of a pancake BEC \cite{pancake}, we take $V_{trap}=\frac{m}{2} \omega
^{2}x^{2}$ where $\omega$ is the harmonic oscillator frequency and in the
transverse direction (i.e. in the direction orthogonal to the $x$-axis) the
size of the condensate is supposed large enough to be considered infinite in
the first approximation. In longitudinal direction the size of the condensate
is of order of magnitude $a=(\hbar/m\omega)^{1/2}$. It is convenient to
introduce some typical reference frequency $\omega_{0}$ so that the trap
frequency is measured in units of $\omega_{0}$, $\omega=\nu\omega_{0}$, $\nu$
being the dimensionless trap frequency.
To make the dynamical equation dimensionless, we introduce
\begin{equation}
\label{subs}\psi(\mathbf{r},t)=\left( 2\sqrt{2}\pi a_{0}^{2} a_{s}\right)
^{-\frac{1}{2}} \exp\left( i\mathbf{k}_{\bot}\mathbf{r}_{\bot}-i\frac{\hbar
k_{\bot}^{2}}{2m}t\right) \Psi(x,t),
\end{equation}
where $\mathbf{r}_{\bot}=(y,z)$ and $a_{0}^{2}=\hbar/m\omega_{0}=a^{2}\nu$,
and make a change of independent variables $x= 2^{-1/4} a_{0} x^{\prime}$, $t=
2^{1/2} t^{\prime}/\omega_{0}$. This results in the canonical form of the NLS
equation with a parabolic potential
\begin{equation}
\label{NLS}i\frac{\partial\Psi}{\partial t}+\frac{\partial^{2}\Psi}%
{\partial{x}^{2}} -2 |\Psi|^{2}\Psi=\frac{1}{2}\nu^{2}x^{2}\Psi,
\end{equation}
where the primes were suppressed. The dimensionless BEC wave function
$\Psi(x,t)$ is normalized according to
\begin{equation}
\label{normaliz}\int_{-\infty}^{\infty} |\Psi(x,t)|^{2}dx=\frac{4\pi}{2^{1/4}}
\frac{\mathcal{N} a_{s}a_{0}}{S},
\end{equation}
where $\mathcal{N}$ is the total number of particles and $S$ is an effective
area of the transverse cross section of the condensate.

A stationary solution $\Psi(x,t)=\Psi(x)\exp(-i\mu t)$, corresponding to the
ground state of BEC, is given by $\Psi(x)$ satisfying the equation
\begin{equation}
\label{NLS-stat}\frac{d^{2}\Psi}{d{x}^{2}}+\mu\Psi-2 |\Psi|^{2}\Psi
=\frac{1}{2}\nu^{2}x^{2}\Psi,
\end{equation}
subject to the zero boundary condition at $|x|\to\infty$ and having no other
zeros. The eigenvalue $\mu$ (chemical potential) is determined by the
normalization (\ref{normaliz}). In dimensionless units the longitudinal size
of the condensate is of order of magnitude $\mu^{1/2}/\nu$ and if $\mu\gg\nu$,
the considerable part of the condensate can be described by TF approximation
in which the term with the second space derivative in (\ref{NLS-stat}) can be
neglected almost everywhere, so that
\begin{equation}
\label{TF}\rho_{0}(x)\equiv|\Psi_{TF}(x)|^{2}=\tfrac12(\mu-\tfrac12\nu
^{2}x^{2})
\end{equation}
with normalization
\begin{equation}
\label{TF0}\int_{-\sqrt{2\mu}/\nu}^{\sqrt{2\mu}/\nu}|\Psi_{TF}(x)|^{2}dx
=\frac{(2\mu)^{3/2}}{3\nu}.
\end{equation}
TF approximation (\ref{TF}) fails at the tails of the density distribution
where the density decays exponentially as $|x|\to\infty$ instead of vanishing
at finite distance $|x|=\sqrt{2\mu}/\nu$ (called TF radius) according to
Eq.~(\ref{TF}). Since only a small part of the condensate's mass is
concentrated in these tails (for example, for $\mu=2$ and $\nu=0.5$, it is
less then $1.5\%$).
Thus in the limit $\mu\gg\nu$ the distribution (\ref{TF}) is assumed to be a
good approximation of the initial density distribution of BEC before switching
off the external potential.

By equating Eqs.~(\ref{normaliz}) and (\ref{TF0}) we determine the value of
the dimensionless chemical potential $\mu$ in terms of experimentally
measurable parameters
\begin{equation}
\label{b3}\mu=\left( 2^{1/4}3\pi\nu\frac{\mathcal{N} a_{s}a_{0}}{S}\right)
^{2/3}.
\end{equation}
Then the condition $\mu\gg\nu$ yields the following criterion
$ \mathcal{N}a_{s} a/S \gg1 $
of applicability of the TF approximation in physical units. In what follows we
deal mainly with two parameters $\nu$ and $\mu$, which completely define the
initial distribution of the condensate density.


After switching off the potential, the condensate evolves according to
Eq.~(\ref{NLS}) with $\nu=0$. If evolution does not lead to wave breaking of
the pulse, then we still can neglect the dispersion effects and use the
hydrodynamic approximation for description of this evolution. As we shall see,
this is the case of initial distribution (\ref{TF}), so that the hydrodynamic
approximation is valid even at asymptotically large values of time considered
in \cite{manakov}. Therefore we pass from the NLS equation (\ref{NLS}) to its
hydrodynamic representation. We represent $\Psi(x,t)$ in the form
\begin{equation}
\label{madelung}\Psi(x,t)=\sqrt{\rho(x,t)}\exp\left( {i}\int^{x} v(x^{\prime
},t)dx^{\prime}\right) ,
\end{equation}
so that substitution of (\ref{madelung}) into (\ref{NLS}) with $\nu=0$ yields
\begin{equation}
\label{hydro}\tfrac12\rho_{t}+(\rho v)_{x}=0,\quad\tfrac12v_{t}+vv_{x}%
+\rho_{x}=0,
\end{equation}
which is subject to the the initial conditions
\begin{equation}
\label{init}\rho(x,0)=\rho_{0}(x),\quad v(x,0)=0,
\end{equation}
where $\rho_{0}(x)$ is taken to be the initial distribution (\ref{TF}). In
(\ref{hydro}) we have neglected the higher space derivatives of the density
$\rho(x,t)$ that correspond to the quantum-pressure contribution.

A problem similar to (\ref{hydro}), (\ref{init}) was studied in nonlinear
optics long ago \cite{Talanov} for the opposite sign of the ``pressure''
$\rho_{x}$ in (\ref{hydro}) which corresponds to evolution of an optical beam
in a focusing Kerr medium. More complex initial conditions were considered in
\cite{ASK66,GS70}, and in the recent paper \cite{SSS02} the same nonlinear
geometrical optics method has been applied to investigation of wave breaking
phenomena in BEC. Here we shall apply the method of \cite{Talanov} to the
problem of BEC expansion.

We look for a solution of (\ref{hydro}), (\ref{init}) in the form
\begin{equation}
\label{anz}\rho(x,t)=\frac{\mu}{2f(t)}\left( 1-\frac{\nu^{2}x^{2}}{2\mu
f(t)^{2}} \right) , \quad v(x,t)=x\phi(t),
\end{equation}
where according to (\ref{init}) the functions $f(t)$ and $\phi(t)$ must
satisfy the initial conditions
\begin{equation}
\label{init2}f(0)=1,\quad\phi(0)=0.
\end{equation}
Substitution of (\ref{anz}) into (\ref{hydro}) gives the relationship between
$f(t)$ and $\phi(t)$,
\begin{equation}
\label{rel}\phi(t)=\frac{f^{\prime}(t)}{2f(t)},
\end{equation}
as well as the differential equation for $f(t)$
\begin{equation}
\label{diff}f^{2}f^{\prime\prime}={2\nu^{2}}.
\end{equation}
This equation can be readily solved with the initial conditions $f(0)=1$,
$f^{\prime}(0)=2f(0)\phi(0)=0$ (see (\ref{init2}) and (\ref{rel})) to give
\begin{equation}
\label{sol}%
\begin{array}
[c]{ll}%
4\nu\, t & =2\sqrt{f(f-1)}\\
& +\ln\left[ 2f-1+ 2\sqrt{f(f-1)}\right] .
\end{array}
\end{equation}
This formula determines implicitly $f$ as a function of $t$ and, hence, the
distribution of density $\rho(x,t)$ according to Eq.~(\ref{anz}). Then the
function $\phi(t)$ is determined by (\ref{rel}), so that distribution of
velocities $v(x,t)=x\phi(t)$ is given by
\begin{equation}
\label{velocity}v(x,t)=\frac{x\nu}{f(t)}\sqrt{1-\frac{1}{f(t)}}.
\end{equation}
Thus, Eqs.~(\ref{anz}), (\ref{rel}), and (\ref{sol}) give the complete
analytic solution of the posed problem in the TF approximation.

One of the effects which accompanies the expansion of the condensate is a
non-monotonic behavior of the current density
\begin{equation}
\label{current}J(x,t)=\rho(x,t) v(x,t).
\end{equation}
Equating its derivatives with respect to $x$ and $t$ to zero yields with the
use of Eqs.~(\ref{hydro}), (\ref{anz}) and (\ref{velocity}) its maximum value
as well as corresponding values of $x$ and $t$:
\begin{align}
\label{Jxt_max} &  J_{m}(x_{m}, t_{m})=\frac{(2\mu)^{3/2}}{27},\nonumber\\
&  x_{m}=\sqrt{\frac{3\mu}{2\nu^{2}}},\quad t_{m}=\frac{\sqrt{3}+\ln
(2+\sqrt{3})}{4\nu},
\end{align}
which with the use of Eq.~(\ref{b3}) can be expressed in terms of
experimentally measurable parameters
\begin{align}
\label{Jxt_max_exp} &  J_{m}(x_{m}, t_{m})=1.17\, \nu\, \frac{\mathcal{N}
a_{s}a_{0}}{S},\nonumber\\
&  x_{m}=2.74\,\nu^{-2/3}\left( \frac{\mathcal{N} a_{s}a_{0}}{S}\right)
^{1/3},\\
&  t_{m}=\frac{\sqrt{3}+\ln(2+\sqrt{3})}{4\nu} \approx0.76\, \nu
^{-1}.\nonumber
\end{align}

The maximum of the density flow is proportional to the number of particles
$\mathcal{N}$, although the time coordinate of the current maximum does not
depend on the density but only on the condensate aspect ratio $\nu$. Let us
consider the typical experiments on the condensate with $\mathcal{N}=10^{5}$
atoms of $^{87}$Rb with scattering length $a_{s}\approx5$nm \cite{Han}. Taking
the length scale of the condensate of order of $a_{0}\approx1\mu$m (what
corresponds to the frequency $\omega_{0}\sim5\cdot10^{3}$Hz) and transversal
radius $\approx10\mu$m one can obtain for $\nu=0.2$ the maximum of the current
density $J_{m}\approx100 $ atoms$\cdot\mu$m$^{-2}\cdot$ms$^{-1}$ in the
coordinate $x_{m}\approx11.5\mu$m (initial TF radius was $\approx13.3\mu$m)
and at the time $t_{m}\approx 1.1\,$ms.

In asymptotic limit of large $t$, when $f(t)\gg f(0)$, we have
\begin{equation}
f(t)\cong2\nu\,t,\quad\phi(t)\cong\frac{1}{2t}.\label{asymp}%
\end{equation}
Hence, the solution (\ref{anz}) takes the form
\begin{equation}
\rho(x,t)\cong\frac{\mu}{4\nu t}\left(  1-\frac{x^{2}}{8\mu t^{2}}\right)
,\quad v(x,t)\cong\frac{x}{2t}.\label{asymp-sol}%
\end{equation}
These formulas describe a hydrodynamic flow ``by inertia'' when the density
becomes so small that the ``pressure'' does not accelerate the gas anymore
(factor $1/2$ in the second formula (\ref{asymp-sol}) corresponds to
definition of the ``time'' variable in Eqs.~(\ref{hydro})).

\begin{figure}[th]
\centerline{\includegraphics[width=6cm,height=5cm,clip]{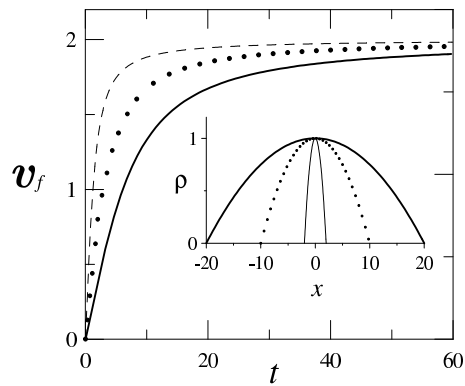}}
\vspace{0.3 true cm}\caption{Time dependence of the front velocity $v_{f}(t)$
for $\mu=2$ and for different $\nu$ (solid line $\nu=0.1$, dotted line
$\nu=0.2$, dashed line $\nu=0.5$). In the insert the corresponding initial
distributions of the condensate density $\rho_{0}(x)$ are depicted. }%
\label{figone}%
\end{figure}

From (\ref{asymp-sol}) we can easily find asymptotic distribution of particles
on their velocities:
\begin{equation}
\label{velo}W(v)dv=\frac{\mu}{2\nu}\left( 1-\frac{v^{2}}{2\mu}\right) dv,
\end{equation}
which gives the number of atoms with velocities in the interval $(v,v+dv)$.
Also, in the framework of the TF approximation Eq.~(\ref{anz}) and with the
use of (\ref{diff}) one obtains the velocity of the condensate front
\begin{equation}
\label{veloc_front}v_{f}(t)=\sqrt{\frac{\mu}{2}}\,\frac{1}{\nu}\frac{d f}{d
t}= \left[ 2\mu\left( 1-\frac{1}{f}\right) \right] ^{1/2}.
\end{equation}
which at $t\to\infty$ asymptotically goes to $v_{f}(\infty)\cong\sqrt{2\mu}$.
The corresponding plots are shown in Fig.~\ref{figone}.

To verify above findings, we have carried out numerical simulations of the
condensate dynamics governed by Eq.~(\ref{NLS}) after switching off the trap
potential, i.e. with $\nu=0$. Initial profile of the condensate was taken as a
numerical solution of Eq.~(\ref{NLS-stat}).

\begin{figure}[th]
\centerline{\includegraphics[width=7cm,height=3cm,clip]{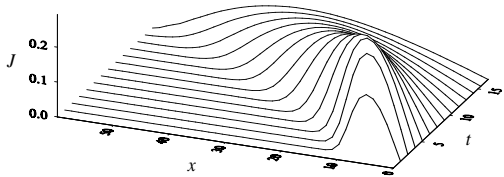}}
\vspace{0.3 true cm}\caption{Spatial-temporal behavior of the density flow
$J(x,t)$ for $\mu=2$ and $\nu=0.2$. }%
\label{figtwo}%
\end{figure}

\begin{figure}[th]
\includegraphics[width=4cm,height=3cm,clip]{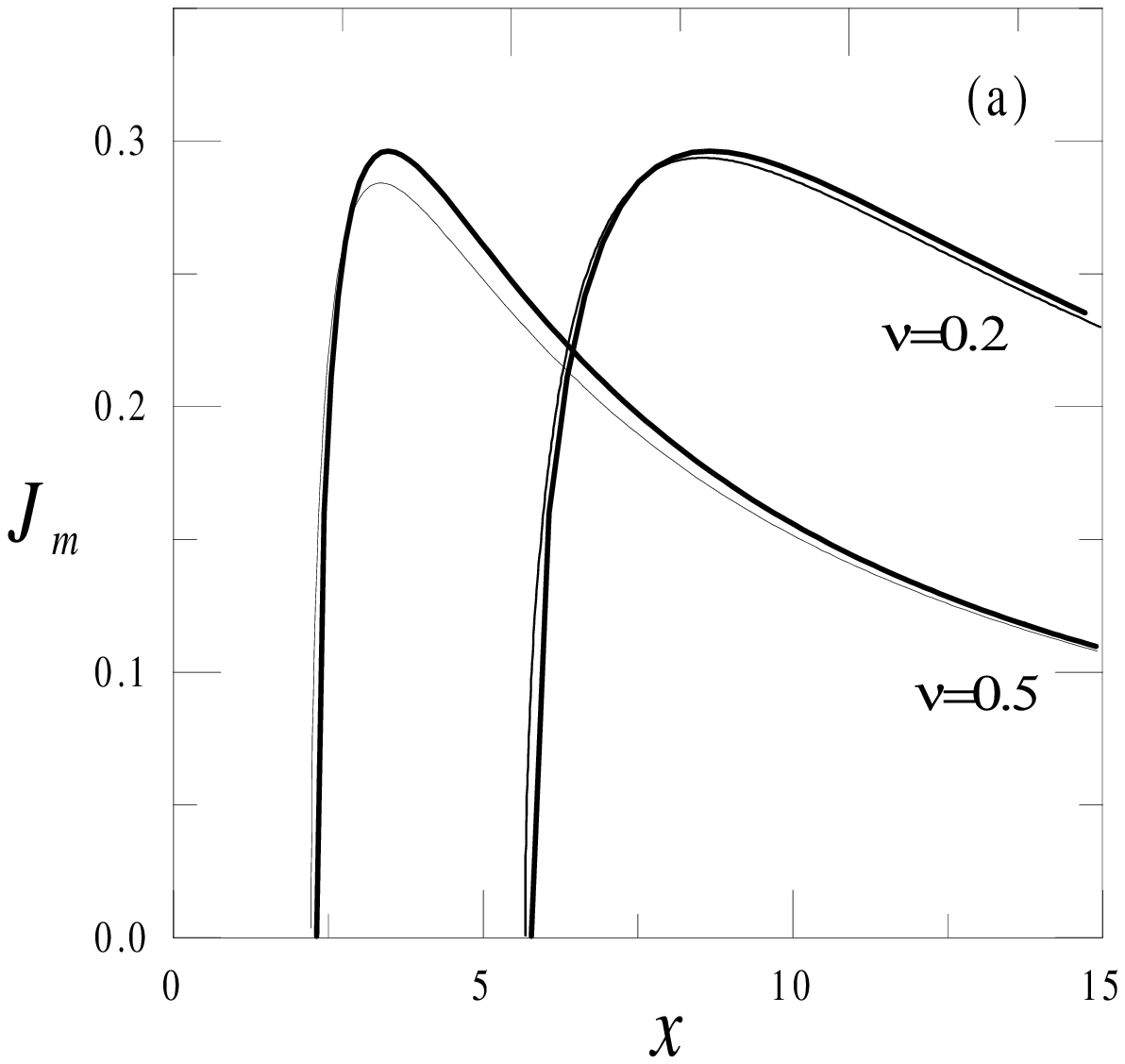}
\includegraphics[width=4cm,height=3cm,clip]{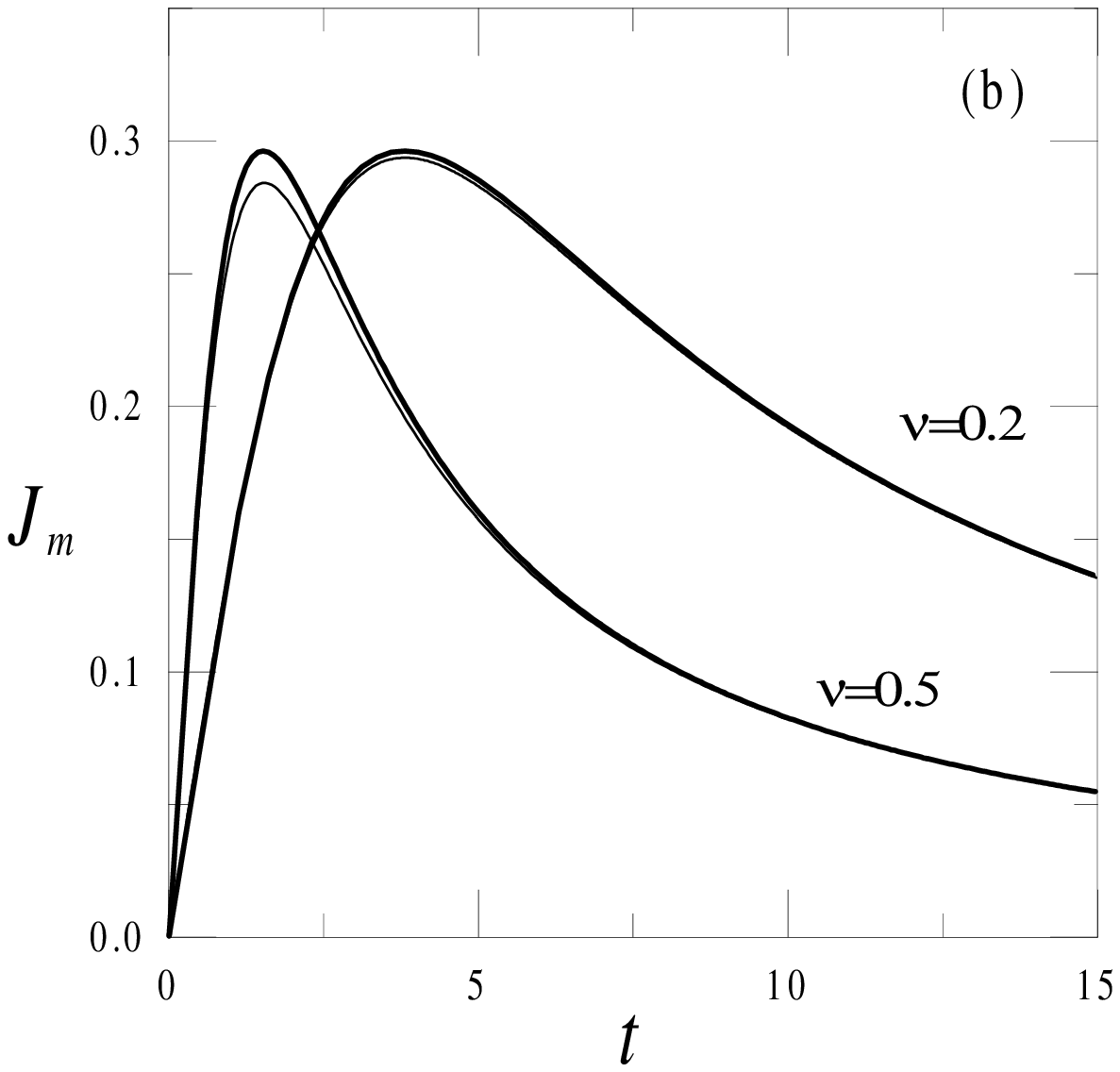}  \vspace{0.3 true
cm}\caption{Dependence of the maximum of the density flow with coordinate (a)
and time (b). Here solid thick lines correspond to the analytical solution and
thin lines show the results of numerical calculations. }%
\label{figthree}%
\end{figure}

In agreement with analytical predictions we found that during expansion the
current density $J(x,t)=\rho(x,t) v(x,t)$ has maximum (see Fig.~\ref{figtwo}).
The space and time coordinates of this maximum which were calculated with the
use of the TF approximation and by direct numerical calculations practically
coincide (see Fig.~\ref{figthree}). The discrepancy between the analytical and
numerical calculations increases with increase of the parameter $\nu$ (see
Fig.~3) what can be explained by lost of the accuracy of the TF approximation
in accordance with the criterion $\mu\gg\nu$.

Let us estimate time when we can consider an expantion of the condensate cloud
only in the axial direction. In the particular case of the pancake geometry
this can be done by conparison of the respect kinetic parts of the initial 3D GP
equation. This gives us a criterium 
$T_{0}/T_{\bot}\approx a^{2}/a_{\bot}^{2}\approx0.01\ll1$ 
where $T_{0}$ and $T_{\bot}$ are characteristics times of
the processes in the axial and radial direction of the condensate
correspondingly. Considering time scale $t\ll T_\bot$ one can neglect the kinetics
of the condensate in the radial direction and
thus consider it to be unchanged in this direction.
In order to estimate $T_0$ and  $T_\bot$  we notice that
velocity has an order of magnitude
$v\sim\sqrt{\mu}$  (see (24)) and thus 
$T_{0}\simeq\frac{a_\bot}{\sqrt{\mu}}$, $a_\bot$ being the transverse radius.
In the case of numerical calculations reported in Figs.1,2 one estimates 
$T_0\sim 0.4$ms (1 dimensional unit) and $T_\bot\sim 40$ms 
(100 dimmensional units) and thus the asymptotic limit described by (21) 
is indeed achieved.

To conclude, it is interesting to make a comparison of our simplified 1D model
with some experiments where the expansion of the condensate cloud was
observed. To follow the real experimental setup we take $\mu=$const what
corresponds to constant density $\rho_{0}$ at the condensate center (see
Eq.~(\ref{TF})) and look for the front velocities $v_{f}(t)$ with different
$\nu$ what corresponds to different sizes (number of particles $\mathcal{N}$)
of the condensate in the transversal directions as was shown in
Fig.\ref{figone}. For larger values $\nu$ (more narrow profile of the
condensate) the front velocity $v_{f}$ is higher than for condensate with
smaller $\nu$. This result of the 1D model is in qualitative agreement with
the experimental observations where faster expansion of the condensate cloud
was observed in the direction with smaller transverse size
\cite{Stamper,Mewes}.

V.V.K. is grateful to A. B. Shvartsburg for valuable discussions. A.M.K. is
grateful to the staff of Centro de F\'{\i}sica da Mat\'eria Condensada,
Universidade de Lisboa, for kind hospitality. The work of V.A.B. has been
supported by the FCT fellowship SFRH/BPD/5632/2001. The work of A.M.K. in
Lisbon has been supported by the Senior NATO Fellowship. A.M.K. thanks also
RFBR (Grant 01--01--00696) for partial support. V.V.K. acknowledges support
from the Programme ``Human Potential-Research Training Networks", contract No. HPRN-CT-2000-00158.

\end{document}